\documentclass[conference]{IEEEtran}
\IEEEoverridecommandlockouts

\usepackage[letterpaper, top=0.75in, bottom=0.75in, left=0.75in, right=0.75in]{geometry}

\usepackage{cite}
\usepackage{amsmath,amssymb,amsfonts}
\usepackage{algorithmic}
\usepackage{graphicx}
\usepackage{textcomp}
\usepackage{xcolor}
\usepackage{algorithm}
\usepackage{algorithmic}
\usepackage{multirow}
\usepackage{subcaption}
\usepackage{bbding}
\usepackage{multicol}
\usepackage{float}
\usepackage{hyperref}

\usepackage{eso-pic} 

\def\BibTeX{{\rm B\kern-.05em{\sc i\kern-.025em b}\kern-.08em
    T\kern-.1667em\lower.7ex\hbox{E}\kern-.125emX}}
\begin{document}

\newcommand{\mycopyrighttext}{
  \footnotesize
  \noindent
  \textcopyright~2025 IEEE. Personal use of this material is permitted. Permission from IEEE must be obtained for all other uses, in any current or future media, including reprinting/republishing this material for
  advertising or promotional purposes, creating new collective works, for resale or redistribution to servers or lists, or reuse of any copyrighted component of this work in other works.\\
  36th IEEE Intelligent Vehicles Symposium (IV 2025) - June 22-25, 2025.
}

\renewcommand{\IEEEtitletopspaceextra}{0.25in}

\title{Evaluation of
Coordination Strategies for Underground Automated Vehicle Fleets
\\ in Mixed Traffic 

\thanks{This work has been supported by Sustainable Underground Mining (SUM) Academy, Project SP-12 2021-2024, and the Industrial Graduate School Collaborative AI \& Robotics funded by the Swedish Knowledge Foundation Dnr:20190128.}
}

\author{
\IEEEauthorblockN{
Olga Mironenko \Envelope}
\IEEEauthorblockA{\textit{Center for Applied Autonomous} \\ \textit{Sensor Systems (AASS)} \\
\textit{Örebro University}\\
Örebro, Sweden \\
olga.mironenko@oru.se}
\and
\IEEEauthorblockN{
Hadi Banaee}
\IEEEauthorblockA{\textit{Center for Applied Autonomous} \\ \textit{Sensor Systems (AASS)} \\
\textit{Örebro University}\\
Örebro, Sweden \\
hadi.banaee@oru.se}
\and
\IEEEauthorblockN{
Amy Loutfi}
\IEEEauthorblockA{\textit{Center for Applied Autonomous} \\ \textit{Sensor Systems (AASS)} \\
\textit{Örebro University}\\
Örebro, Sweden \\
amy.loutfi@oru.se}
}

\AddToShipoutPictureBG*{
  \AtPageUpperLeft{
    \hspace{0.8in}
    \raisebox{-3.0\baselineskip}[0pt][0pt]{
      \parbox{\textwidth}{\raggedright \mycopyrighttext}
    }
  }
}

\maketitle

\begin{abstract}
This study investigates the efficiency and safety outcomes of implementing different adaptive coordination models for automated vehicle (AV) fleets, managed by a centralized coordinator that dynamically responds to human-controlled vehicle behavior.
The simulated scenarios replicate an underground mining environment characterized by narrow tunnels with limited connectivity. 
To address the unique challenges of such settings, we propose a novel metric — Path Overlap Density (POD) — to predict efficiency and potentially the safety performance of AV fleets. 
The study also explores the impact of map features on AV fleets performance.
The results demonstrate that both AV fleet coordination strategies and underground tunnel network characteristics significantly influence overall system performance. While map features are critical for optimizing efficiency, adaptive coordination strategies are essential for ensuring safe operations.
\end{abstract}


\section{Introduction}
In underground mining operations, vehicles must navigate confined spaces with narrow tunnels, where passing opportunities are often limited due to vehicle size and spatial constraints. A typical underground mine consists of a network of tunnels, usually connected by a single main tunnel that serves as the primary access route. 
Within this environment, partially automated vehicles perform repetitive tasks, such as transporting ore from draw points (DPs), where ore is extracted, to designated ore passes (OPs) along predefined routes.  

However, these predefined routes often intersect, creating path overlaps while vehicles navigate shared sections of the tunnel network. Since vehicles repeatedly traverse the same paths, interactions at these overlapping sections occur consistently, presenting operational challenges. These challenges become even more complex when manually controlled vehicles (MVs) enter the area, as human driver behavior introduces additional unpredictability. To address this issue, adaptive coordination models that can dynamically respond to human-driven vehicle movements in real time are required \cite{Mironenko}.  

To effectively evaluate the interactions between automated vehicle (AV) fleets and MVs in complex underground mining environments, it is crucial to test various coordination strategies. However, due to varying traffic volumes, tunnel layouts, and dynamic vehicle interactions, a more sophisticated analytical approach is required to assess the key factors influencing traffic performance. The impact of these factors depends on the specific conditions within the environment.  

This study demonstrates that both AV fleet coordination strategies and tunnel network characteristics play a significant role in overall system performance. While map features are critical for optimizing efficiency, effective coordination strategies are essential for ensuring safe operations and mitigating potential disruptions.

\section{Related Work}

In traffic flow theory, traffic density, volume, and intensity metrics are commonly defined as the number of vehicles per unit length of a roadway (e.g., vehicles per kilometer or vehicles per kilometer per lane), within a given area, or passing a specific point on the road over a given time period \cite{Darwish}, \cite{Gold}. Vehicle-specific traffic impact metrics characterize how a particular vehicle is affected by surrounding traffic conditions. These metrics include various measures such as time headway (the time gap between two consecutive vehicles) \cite{Winsum}, following distance, and time-to-collision \cite{Lee}. Road network density, as measured through street connectivity, and its impact on traffic flow are evaluated in research using traffic simulations. Good street connectivity refers to streets that have multiple links and intersections, while poor connectivity is characterized by dead ends \cite{Zlatkovic}. Road networks are often modeled as graphs, and the concept of overlapping paths in a route is widely studied in graph theory and graph traversal algorithms, primarily in the context of pathfinding, routing efficiency, traffic management, and robot navigation. In this context, overlapping or intersecting paths refer to routes in a graph that share one or more edges or vertices \cite{Golumbic}. However, these traditional approaches primarily focus on high-connectivity networks, where the objective is to identify multiple routes between two nodes, find vertex and edge-disjoint paths \cite{Tzoreff, Omran}, or determine the shortest obstacle-avoiding path \cite{Mitchell} to improve efficiency and reduce congestion.

In the context of AVs coordination, two main approaches emerge: centralized and decentralized \cite{Cao}. Within environments characterized by spatial constraints and low connectivity, the centralized approach proves advantageous for coordinating vehicle interactions on shared routes and intersecting trajectories, thereby preventing conflicts. Achieving this level of coordination is challenging with decentralized approaches, where individual AVs make autonomous decisions \cite{Ismail, Yan}. Such autonomy can lead to conflicts on overlapping paths, traffic congestion, or collisions. However, adapting the coordination strategies to accommodate the unpredictable behavior of an MV presents challenges for both centralized and decentralized approaches.

\section{Methodology}

The purpose of the experiments is to evaluate the impact of different coordination strategies and map features on the performance of AV fleets in mixed traffic scenarios that involve fleets of AVs and an MV. To achieve this, a methodological framework was developed. This framework consists of simulation components, including Map Generator, ORU Coordination Framework, and analysis modules.

\subsection{Map Generator} 

A map generation tool was developed and utilized to create maps that reflect real-world underground mine layouts and follow underground mining traffic constraints. For experimental purposes, we also tested a set of maps featuring higher connectivity than traditional underground mine layouts to explore scenarios beyond the typical low-connectivity environments. 

\begin{figure}[tb]
    \centering
    \begin{subfigure}[b]{0.49\columnwidth}
        \centering
        \includegraphics[width=\textwidth]{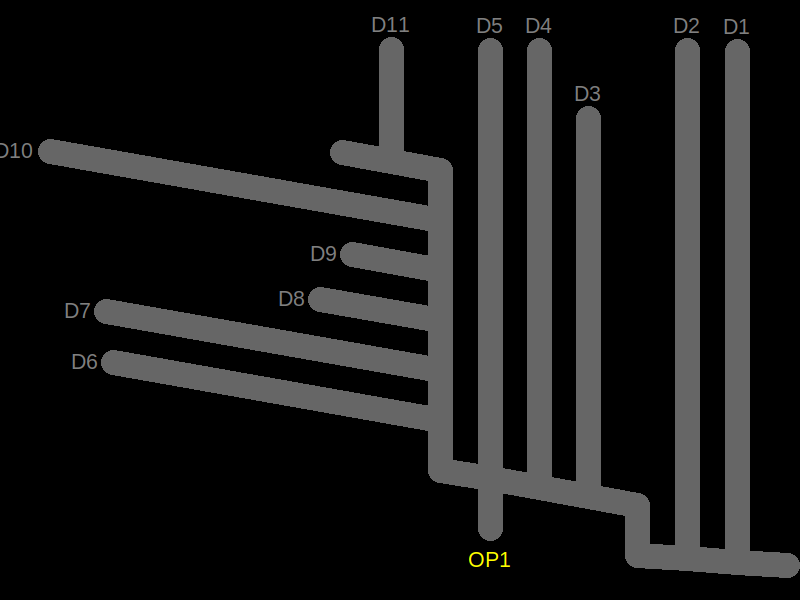}
        \caption{Map 3 (with one OP)}
        \label{map3}
    \end{subfigure}
    \hfill
    \begin{subfigure}[b]{0.49\columnwidth}
        \centering
        \includegraphics[width=\textwidth]{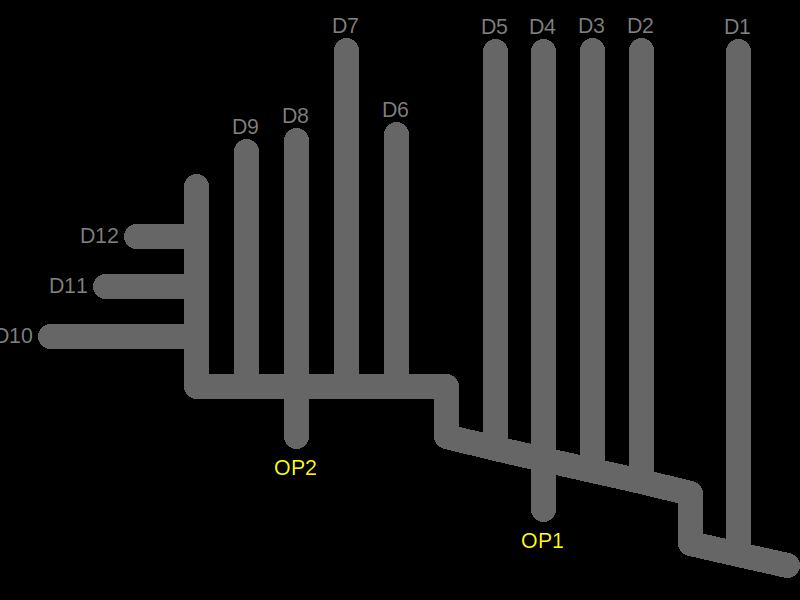}
        \caption{Map 6 (with two OPs)}
        \label{map6}
    \end{subfigure}
    \caption{Examples of generated maps with low connectivity, with (a) one and (b) two ore passes (OPs).}
    \label{fig:maps}
\end{figure}

\begin{figure}[tb]
\centering
\begin{minipage}{0.49\columnwidth}
    \centering
   \includegraphics[width=\textwidth]{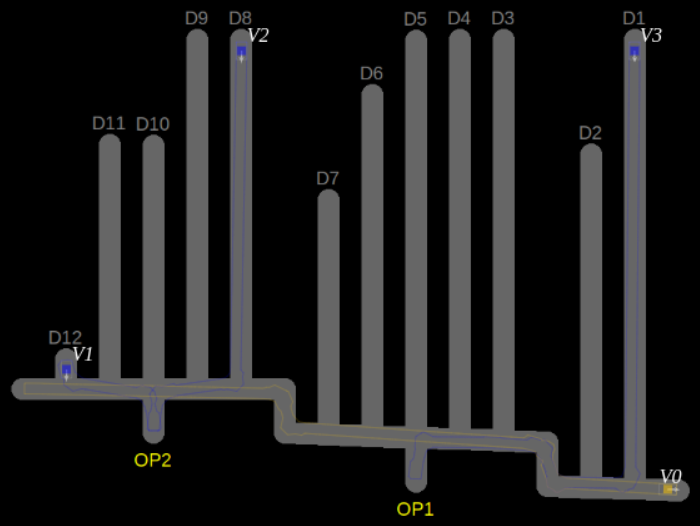} 
    \caption{Map 1. Low connectivity}
    \label{low}
\end{minipage}
\hfill
\begin{minipage}{0.49\columnwidth}
    \centering    
    \includegraphics[width=\textwidth]{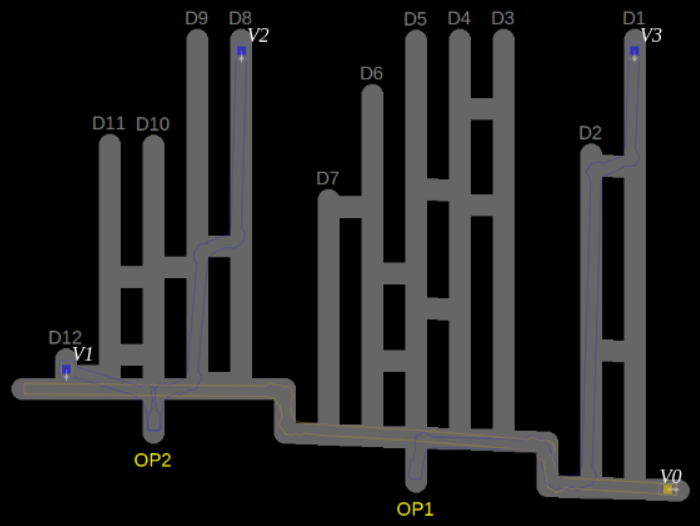} 
    \caption{Map 1. High connectivity}
    \label{high}
\end{minipage}
\end{figure}

Ten distinct maps were created with either one OP (Fig.~\ref{map3}) or two OPs (Fig.~\ref{map6}), each with two connectivity levels: low and high. In the low-connectivity variant, all tunnels are dead-end and connected exclusively through a main tunnel, limiting alternative routes (Fig.~\ref{low}). In contrast, the high-connectivity variant includes additional connections between tunnels, providing more route options (Fig.~\ref{high}). Each tunnel containing a draw point (DP) was assigned a unique identifier that serves as the starting location for placing AVs. In maps with two OPs, the nearest OP was selected as the destination point during pathfinding for AVs (Fig.~\ref{low} and Fig.~\ref{high}). Ten distinct AV starting position configurations were generated randomly and consistently applied across both map connectivity variants to ensure a fair comparison.

To maintain a controlled study, we limit our analysis to two map features: connectivity levels and the number of OPs, as these were the primary parameters in map generation.

\subsection{ORU Coordination Framework} 

The generated maps are used to design and conduct experiments within the ORU Coordination Framework\footnote{\url{https://github.com/FedericoPecora/coordination_oru}}, a platform for centralized multi-robot motion planning, coordination, and control. The interactions within the AV fleet and its coordination methods are outlined in \cite{Pecora2012,Pecora2018}. This framework has been modified to integrate an interactively human-operated vehicle, controlled via a mouse and keyboard. The operator can select destinations, control speed, and accelerate or decelerate, stopping whenever desired. Furthermore, for the purpose of running simulated experiments, the framework has been adapted to simulate such human driver behavior to enable seamless interaction with centrally coordinated AV fleets.

\textbf{Coordination Strategies:} 
The initial implementation of AV fleets' coordination within the ORU Coordination Framework utilizes the ‘Closest-first' heuristic. According to this approach, if all AVs are regarded as equal in rights of way, the vehicle nearest to the intersection has precedence. Alternative heuristics may assign priority based on attributes such as vehicle ID, color, or size. To evaluate the effectiveness of coordination strategies for AV fleets in mixed traffic conditions, we have introduced an additional rule ‘AV-first’ and several coordination strategies specifically designed to manage AV fleets in response to the behavior of MVs. The ‘AV-first’ rule grants AVs the priority to proceed before MVs at intersections. 

\textit{Dynamic Change of Priorities.} This strategy is applied when an MV violates a precedence traffic rule (AV-first). In such cases, priority at crossroads is temporarily reassigned to the MV by the affected AVs. The AVs continue driving but decelerate as necessary, ensuring they have sufficient time to stop before reaching the crossroad, allowing the MV to pass first.

\textit{Stops.} In similar precedence violation scenarios, this strategy involves the affected AVs stopping as soon as possible, even at a considerable distance from the crossroad. The AVs remain stationary until the MV has safely passed the crossroad, after which they resume driving.

\textit{Rerouting.} When an MV travels significantly slower than the average traffic speed, the AVs are coordinated to reroute whenever feasible. Although in scenarios where the MV does not violate the precedence rule and AVs are not required to stop because of the MV, the slow-moving MV can still affect the system's efficiency.

\subsection{Measuring System's Performance} 

The system's performance is quantified through two metrics: \textit{Efficiency} is measured by the number of missions completed by AVs, and \textit{Safety} is measured by the collision rate, defined as the ratio of the number of collisions to the number of priority rule violations committed by an MV. Collisions may occur when an AV does not have sufficient time to decelerate and stop before reaching an intersection, potentially colliding with the MV. The impact of the coordination strategies employed for AV fleets and map features is analyzed using these metrics. 

A new metric, \textbf{Path Overlap Density (POD)}, is defined to quantify vehicle interactions within shared segments of paths.
Unlike traffic flow theory-based metrics discussed in related work, POD quantifies the interactions between vehicles when their predefined trajectories overlap along a road segment or intersect at a crossroad. Specifically, for each vehicle, POD is calculated by identifying the path segments that overlap with those of other vehicles.
Each overlapping segment, a part of the entire path vector, is represented as a vector of poses with associated coordinates. The length of this overlapping segment is then divided by the total length of the other vehicle's path. 
This ratio indicates how one vehicle’s movement influences another’s trajectory within the shared segment.

If the same poses belong to more overlapping segments within other vehicles’ paths, this part of the path is considered denser, thereby increasing the path overlap density for the vehicle traversing it. The POD for each point along a vehicle’s path is computed as the sum of these ratios:

\begin{equation}
\text{POD}_i(x) = \sum_{j \ne i} \frac{\text{POL}_{ij}(x)}{\mathrm{len}(\mathrm{path}_j)}
\end{equation}

where:
\begin{itemize}
    \item \( \text{POL}_{ij}(x) \), Path Overlap Length, is the length of the segment where the paths of vehicles \( i \) and \( j \) overlap, and which contains the pose \( x \),
    \item \( \mathrm{len}(\mathrm{path}_j) \) is the total path length of vehicle \( j \),
    \item The summation considers all vehicles \( j \neq i \) whose paths overlap with that of vehicle \( i \).
\end{itemize}

\begin{figure}[tb]
    \centering
    \begin{subfigure}{0.29\columnwidth}
        \centering
        \includegraphics[width=0.895\textwidth]{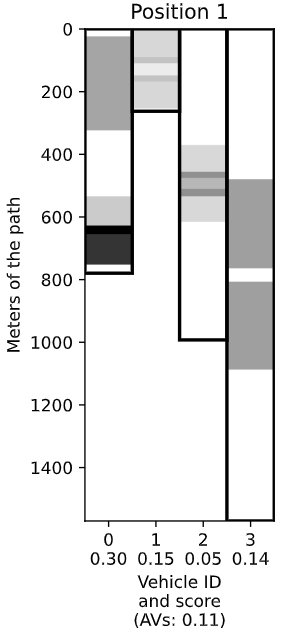}
        \caption{Individual scores per path. Position 1.}
        \label{fig:pod_individual}
    \end{subfigure}
    \hfill
    \begin{subfigure}{0.67\columnwidth}
        \centering
        \includegraphics[width=1\textwidth]{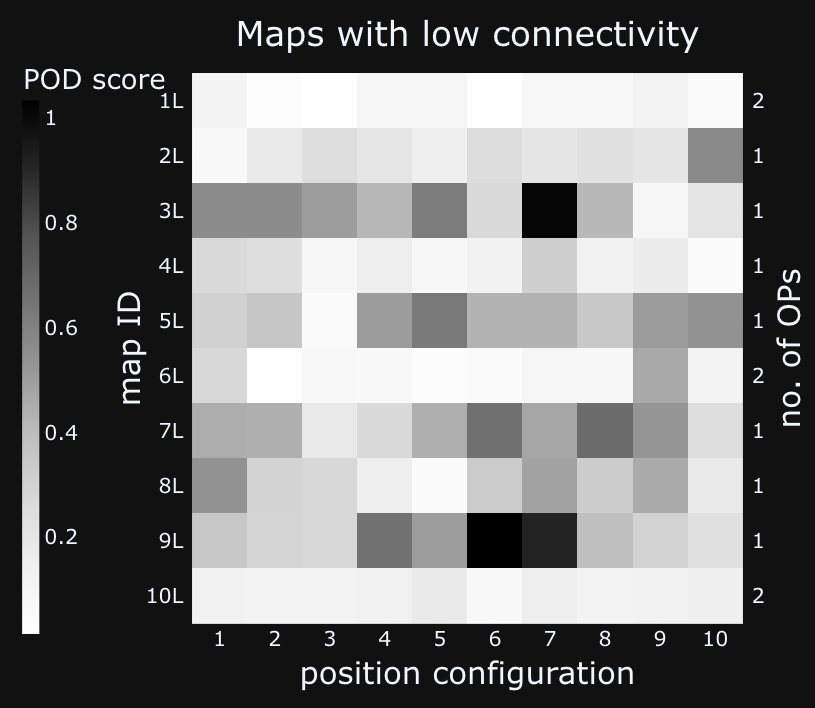}
        \caption{Average scores for all AVs' paths per map and position.}
        \label{fig:pod_scores}
    \end{subfigure}
    \caption{Path Overlap Density (POD) scores for vehicles' paths in maps with low connectivity.}
\end{figure}

To calculate the POD score for each vehicle's path, we sum the PODs along the path and divide by the total path length. An example from Map 1 with low connectivity and Position configuration 1, depicting the path overlap density for four vehicles' paths, is visualized in Fig.~\ref{fig:pod_individual}. Each vehicle is identified by its corresponding ID, and the POD score for each path is displayed under the ID. The length of each path is indicated along the axis. To further assess system performance, we calculate the average POD score for all AVs' paths in a scenario. This is achieved by summing the overall POD scores for each point along all AVs' paths and then dividing by the total length of these paths. This score for AV1-AV3 is displayed in brackets in the same figure (V0, in this example, is an MV).

An example of POD scores across different maps with low connectivity, presented as a matrix and, for simplicity, visualized using a 10x10 heatmap, is shown in Fig.~\ref{fig:pod_scores}. Such heatmaps represent the 10 map configurations and 10 position configurations for each map connectivity scenario, which are indicated by 'L' (low) or 'H' (high) next to the map number. The number of OPs is marked along the second axis.

POD serves as an environment-based feature, quantified after the vehicle paths are generated by a path planner. It is calculated offline before running simulated scenarios.
For the experiments, paths are generated using the Probabilistic Roadmap (PRM) algorithm \cite{Kavraki} from the Open Motion Planning Library (OMPL)
\footnote{\url{https://ompl.kavrakilab.org/planners.html}} \cite{Sucan}, which identifies the shortest optimal routes. These routes are then utilized by the central coordinator of the ORU Coordination Framework to ensure efficient route selection for the experiments.

POD is particularly well-suited for scenarios involving predefined paths that are repeatedly traversed on a known map. To evaluate the effectiveness of the metric, empirical validations are conducted, alongside an analysis of statistical correlations. By correlating POD scores with system performance metrics (i.e., the number of missions completed and the collision rate), we aim to understand the relationship between these variables.

We use the following analytical methodology in this study. To evaluate the overall system performance, we aggregate the results for all AVs in each simulation. After executing the scenarios and quantifying the efficiency and safety outcomes across 10 map configurations and 10 position configurations for each coordination strategy, we correlate these metrics with the POD scores. We use scatter plots to visualize the number of completed missions and collision rates. To analyze the relationship between efficiency and safety outcomes with POD scores, and to assess the impact of coordination strategies across different scenarios, we employ trendlines for each strategy within these plots. We then compare these results to baseline conditions where no coordination strategies were applied to AV fleets in mixed traffic, to evaluate the impact of different coordination strategies and map features on system performance.

\section{Experimental Setup}

\subsection{Traffic Regulations}

The framework employs predefined traffic rules to manage interactions between AVs and the MV. These rules include the ‘AV-first’ and ‘Closest-first’ prioritization protocols, consistently applied across all simulation scenarios. The ‘Closest-first’ rule is applied exclusively among AVs. In some scenarios, the MV consistently violates the ‘AV-first’ rule, while AVs adhere to the ‘Closest-first’ rule without exceptions. In other scenarios, the MV violates the speed rule by driving significantly slower than the average traffic.

The number of vehicles is kept consistent across all scenarios, with three AVs (blue), identified as V1, V2, and V3, and one MV (yellow) identified as V0 (see Map 1 in Figures~\ref{low} and \ref{high}). The MV travels exclusively along the main tunnel, moving from right to left and back. The AVs move along the paths connecting DPs and OPs, completing trips to and from these points as a single mission, according to the position configurations described earlier. Their respective trajectories are visualized.

All vehicles are modeled as trucks and follow a predefined speed rule with the maximum achievable speed for a vehicle set at 5.6 {\footnotesize$m/s$}, which is established as the speed limit in the experimental scenarios. If sufficient distance is available and no obstacles are present along the route, vehicles aim to achieve their maximum speed. In scenarios where the MV violates the speed rule, it travels at a reduced speed significantly lower than the traffic flow - 1.0 {\footnotesize$m/s$}. Acceleration and deceleration parameters, linked to the maximum speed, are determined based on the characteristics of different vehicle types as outlined in \cite{Bokare}. All vehicles share an acceleration rate of 0.3 {\footnotesize$m/s^2$} and a deceleration rate of 0.5 {\footnotesize$m/s^2$}. Vehicles maintain a minimum safety distance from their front bumper to the preceding vehicle’s safety distance, which also includes the required proximity to intersections. For the experiments, the safety distance is set to 5 meters. In simulations, it is visually represented as a white frame around vehicles.

\subsection{Design of Scenarios}

The scenarios are developed around two main cases of traffic rule violations resulting from human behavior: \textit{Priority rule violation} and \textit{Speed rule violation}. A third case involves violations of both traffic rules. To establish a reference point for evaluating the impact of various coordination strategies, the \textbf{Baseline Scenarios for Mixed Traffic} were introduced for each of the three cases. These scenarios illustrate traffic conditions where no coordination strategies are applied to AV fleets in response to the MV violating traffic rules. The baselines serve as a control for comparative analysis against the outcomes of the coordination strategies adapted to the behavior of the MV. 

\textbf{Case I - Priority Rule Violation:} involves an MV that disregards priority rules at intersections and does not yield the right of way as required by the traffic control system. In Scenario 1.1, the coordination strategy involves dynamically adjusting the priority of the affected AVs, granting immediate priority to the violating MV once the system detects the infraction - specifically 30 meters before the intersection if the MV's speed indicates it is not slowing down and cannot stop in time. In Scenario 1.2, the strategy involves stopping the affected vehicles under the same conditions of violation.

\textbf{Case II - Speed Rule Violation:} examines violations of speed regulations, where the MV operates at a speed significantly lower than the average traffic flow, potentially causing disruptions to the overall traffic system. The tested coordination strategy in this scenario involves rerouting traffic to accommodate the slower-moving vehicle. However, rerouting is tested only on maps with high connectivity, where alternative routes are available to accommodate deviations. 

\textbf{Case III - Speed Rule and Priority Rule Violations:} addresses combined violations, where the MV breaches both priority and speed regulations. The coordination strategies mentioned earlier are evaluated across four sub-cases to handle the compounded risks. In Scenario 3.1, dynamic change of priorities is tested. In Scenario 3.2, AVs stops are employed to manage the violations. Scenario 3.3 applies a combination of rerouting and dynamic priority changes, while Scenario 3.4 combines rerouting with AVs stops.

The details of the performed simulations are summarized in Table~\ref{scenarios_summary}. Each scenario highlights the MV's violations and the corresponding coordination strategies. A total of 1700 simulations were conducted, with each simulation lasting 30 minutes.

\begin{table*}[tb]
\centering
\caption{Summary of Scenarios}
\label{scenarios_summary}
\renewcommand{\arraystretch}{1.3}  
\resizebox{1.0\textwidth}{!}{  
\begin{tabular}{|l|l|l|cc|c|}
\hline
\multirow{2}{*}{\textbf{Scenario}} & \multirow{2}{*}{\textbf{MV's Violation}} & \multirow{2}{*}{\textbf{AV Fleet Coordination Strategy}} & \multicolumn{2}{c|}{\textbf{Map Connectivity}} & \multirow{2}{*}{\textbf{No. simulations}} \\ \cline{4-5}
 &  &  & \textbf{Low} & \textbf{High} &  \\ \hline
\hline

\hline
\multicolumn{6}{|c|}{\textbf{Case I - Priority Rule Violation}} \\ \hline
\textbf{Baseline 1} & Priority rule violation & No strategy & \checkmark & \checkmark & 200 \\ \hline
Scenario 1.1 & Priority rule violation & Dynamic change of priorities & \checkmark & \checkmark & 200 \\ \hline
Scenario 1.2 & Priority rule violation & Stops & \checkmark & \checkmark & 200 \\ \hline
\hline

\hline
\multicolumn{6}{|c|}{\textbf{Case II - Speed Rule Violation}} \\ \hline
\textbf{Baseline 2} & Speed rule violation & No strategy & \checkmark & \checkmark & 200 \\ \hline
Scenario 2 & Speed rule violation & Rerouting & & \checkmark & 100 \\ \hline
\hline

\hline
\multicolumn{6}{|c|}{\textbf{Case III - Speed Rule and Priority Rule Violations}} \\ \hline
\textbf{Baseline 3} & Speed rule and Priority rule violations & No strategy & \checkmark & \checkmark & 200 \\ \hline
Scenario 3.1 & Speed rule and Priority rule violations & Dynamic change of priorities & \checkmark & \checkmark & 200 \\ \hline
Scenario 3.2 & Speed rule and Priority rule violations & Stops & \checkmark & \checkmark & 200 \\ \hline
Scenario 3.3 & Speed rule and Priority rule violations & Rerouting and Dynamic change of priorities & & \checkmark & 100 \\ \hline
Scenario 3.4 & Speed rule and Priority rule violations & Rerouting and Stops & & \checkmark & 100 \\ \hline
\hline
\multicolumn{5}{|r|}{\textbf{Total no. simulations}} & \textbf{1700} \\ \hline
\end{tabular}
}
\end{table*}

\section{Results}

Baseline scenarios in Cases I and III demonstrate maximum efficiency, measured by the number of missions completed, but at the cost of safety, with the highest collision rates. Despite MVs' priority rule violations, AVs maintain their operation uninterrupted, without stopping, achieving maximum possible efficiency. However, this unmodified behavior of AVs significantly compromises safety, leading to the highest collision rates in the Baselines. These results are consistent across maps with low and high connectivity. 

Unlike Cases I and III, Baseline 2 for Case II shows the MV that adheres to the priority rule but breaches the speed rule. The observed behavior forces AVs to trail behind the slower MV which affects only efficiency. The results for this scenario indicate the lowest efficiency levels observed. 

The results of the Spearman Rank statistical correlation with the corresponding POD scores for all scenarios are presented in Table~\ref{stat_cor_missions}.
A strong negative correlation, with coefficients ranging from -0.830 to -0.762, was observed between POD scores and the number of missions completed. All correlations were statistically significant, with $p$-values ranging from $7.3 \times 10^{-26}$ to $3.6 \times 10^{-20}$. 
Similar correlation strengths were found across maps with both low and high connectivity.
Additionally, positive correlations ranging from significantly strong to moderate were found between POD scores and the collision rate. These correlations were also statistically significant, with $p$-values ranging from $6.3 \times 10^{-29}$ to $6.8 \times 10^{-3}$.
These values indicate that lower POD scores are associated with higher efficiency and enhanced safety, demonstrating the relationship between the proposed metric, Path Overlap Density (POD), and the key performance outcomes.

\begin{table*}[tb]
\centering
\caption{Spearman Rank Correlations of POD Scores with Number of Completed Missions and Collision Rate}
\label{stat_cor_missions}
\renewcommand{\arraystretch}{1.3}
\resizebox{1.0\textwidth}{!}{
\begin{tabular}{|l|l|c|c|c|c|}  
\hline
\multirow{2}{*}{\textbf{Scenario}} & \multirow{2}{*}{\textbf{AV Fleet Coordination Strategy}} & \multicolumn{2}{c|}{\textbf{Number of Completed Missions}} & \multicolumn{2}{c|}{\textbf{Collision Rate}} \\  
\cline{3-6}
 & & \textbf{Low Connectivity} & \textbf{High Connectivity} & \textbf{Low Connectivity} & \textbf{High Connectivity} \\  \hline
\hline

\hline
\multicolumn{6}{|c|}{\textbf{Case I - Priority Rule Violation}} \\ \hline 
\textbf{Baseline 1} & \textbf{No strategy} & -0.776 & -0.771 & 0.849 & 0.554 \\  
\hline
Scenario 1.1 & Dynamic change of priorities & -0.824 & -0.816 & 0.737 & 0.433 \\  
\hline
Scenario 1.2 & Stops & -0.827 & -0.805 & 0.810 & 0.519 \\  \hline
\hline

\hline
\multicolumn{6}{|c|}{\textbf{Case II - Speed Rule Violation}} \\ \hline  
\textbf{Baseline 2} & \textbf{No strategy} & -0.830 & -0.769 &  &  \\  
\hline
Scenario 2 & Rerouting &  & -0.773 &  &  \\  \hline
\hline

\hline
\multicolumn{6}{|c|}{\textbf{Case III - Speed Rule and Priority Rule Violations}} \\ \hline  
\textbf{Baseline 3} & \textbf{No strategy} & -0.776 & -0.771 & 0.793 & 0.740 \\
\hline
Scenario 3.1 & Dynamic change of priorities & -0.781 & -0.781 & 0.436 & 0.282 \\
\hline
Scenario 3.2 & Stops & -0.804 & -0.762 & 0.269 & 0.300 \\
\hline
Scenario 3.3 & Rerouting and Dynamic change of priorities &  & -0.776 &  & 0.380 \\
\hline
Scenario 3.4 & Rerouting and Stops &  & -0.772 &  & 0.356 \\
\hline
\end{tabular}
}
\end{table*}

\begin{figure*}[tb]
    \centering
    \begin{subfigure}{0.396\textwidth}
        \centering
        \includegraphics[width=1.05\textwidth]{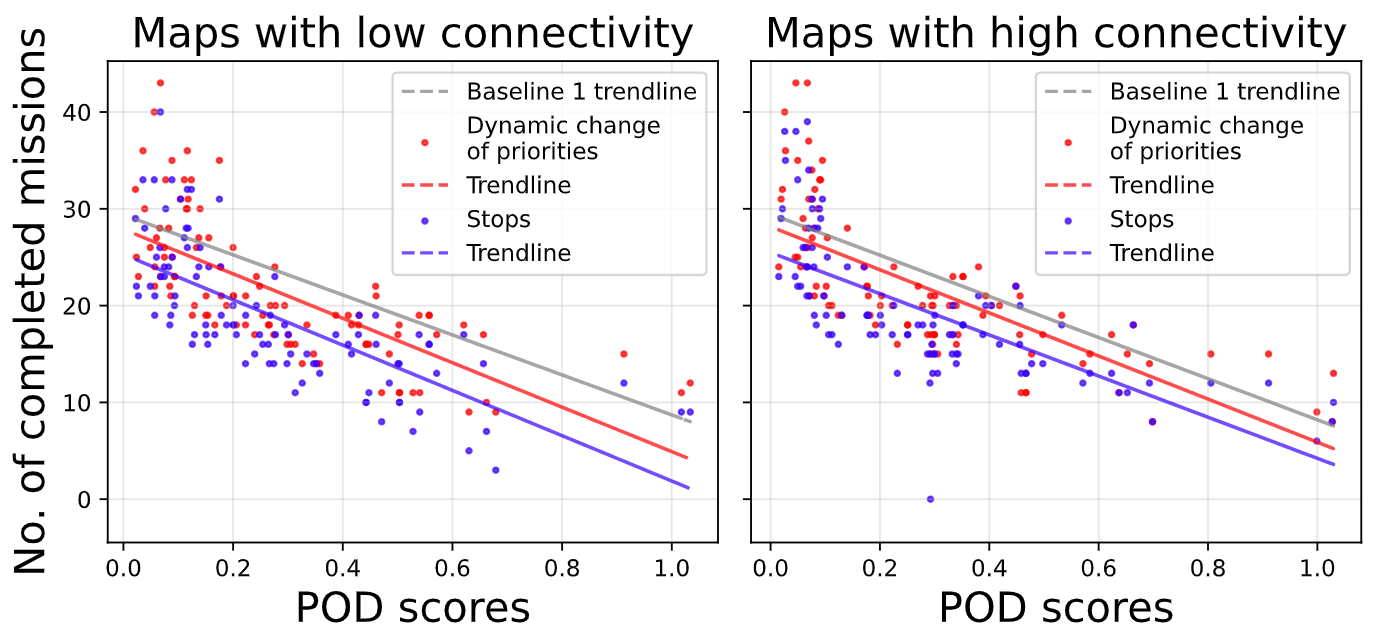}
        \caption{Case I. Scenarios 1.1 versus Scenario 1.2 across two levels of map connectivity.}
        \label{fig:strategy_comparison_1}
    \end{subfigure}
    \hfill
    \begin{subfigure}{0.37\textwidth}
        \centering
        \includegraphics[width=1.05\textwidth]{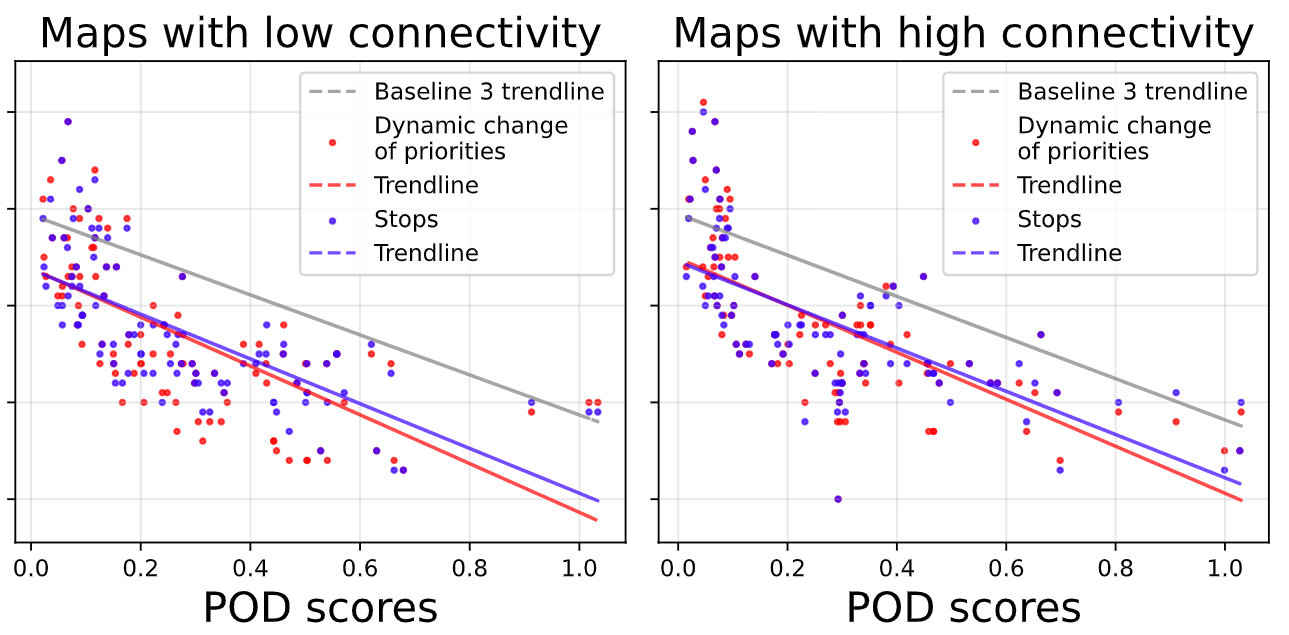}
        \caption{Case III. Scenario 3.1 versus Scenario 3.2 across two levels of map connectivity.}
        \label{fig:strategy_comparison_31}
    \end{subfigure}
    \hfill
    \begin{subfigure}{0.1915\textwidth}
        \centering
        \includegraphics[width=1\textwidth]{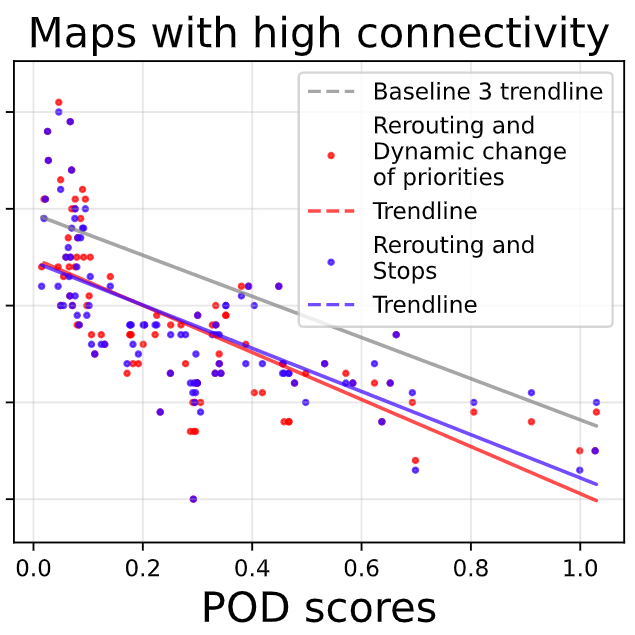}
        \caption{Case III. Scenario 3.1 versus Scenario 3.2.}
        \label{fig:strategy_comparison_32}
    \end{subfigure}
    \caption{Strategies comparison in terms of efficiency (number of completed missions).}
    \label{fig:str_cmp_efficiency}
\end{figure*}

Strong and similar correlations in efficiency across all scenarios result in parallel trendlines with varying intercepts, with optimal outcomes observed in Baselines 1 and 3. 
These patterns suggest that deviations from the Baselines' trendlines in various scenarios may serve as indicators of coordination strategy effectiveness (Fig.~\ref{fig:str_cmp_efficiency}).
Results closer to Baseline trendlines indicate more effective coordination strategies. The trendline with the highest intercept on the Y-axis (i.e., number of completed missions) for the same POD scores will indicate the most effective strategy in terms of efficiency under consistent conditions.

\begin{figure*}[tb]
    \centering
    \begin{subfigure}{0.3965\textwidth}
        \centering
        \includegraphics[width=1.05\textwidth]{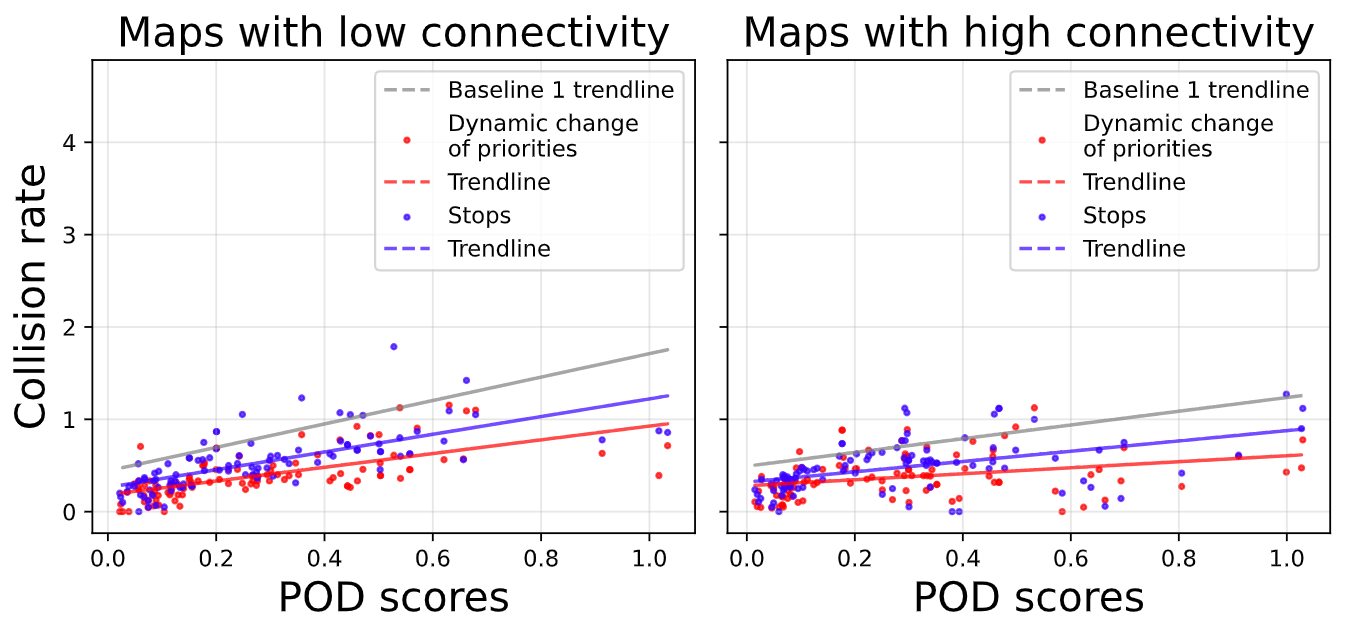}
        \caption{Case I. Scenarios 1.1 versus Scenario 1.2 across two levels of map connectivity.}
        \label{fig:strategy_comparison_1_col}
    \end{subfigure}
    \hfill
    \begin{subfigure}{0.373\textwidth}
        \centering
        \includegraphics[width=1.05\textwidth]{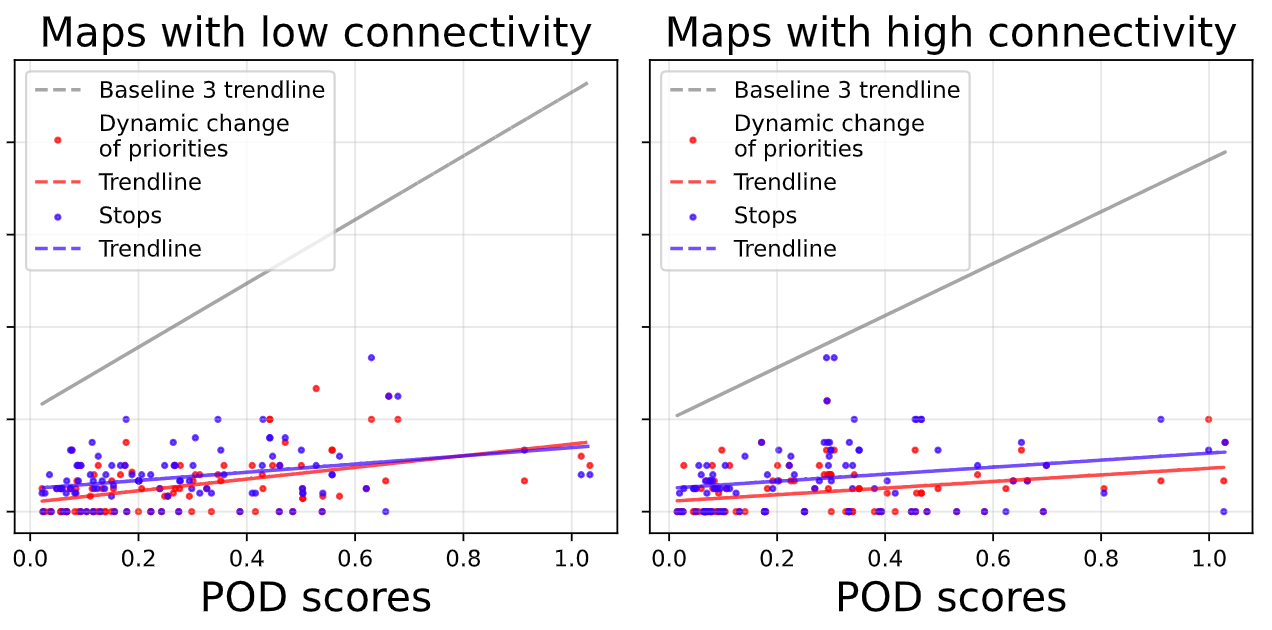}
        \caption{Case III. Scenario 3.1 versus Scenario 3.2 across two levels of map connectivity.}
        \label{fig:strategy_comparison_31_col}
    \end{subfigure}
    \hfill
    \begin{subfigure}{0.1935\textwidth}
        \centering
        \includegraphics[width=1\textwidth]{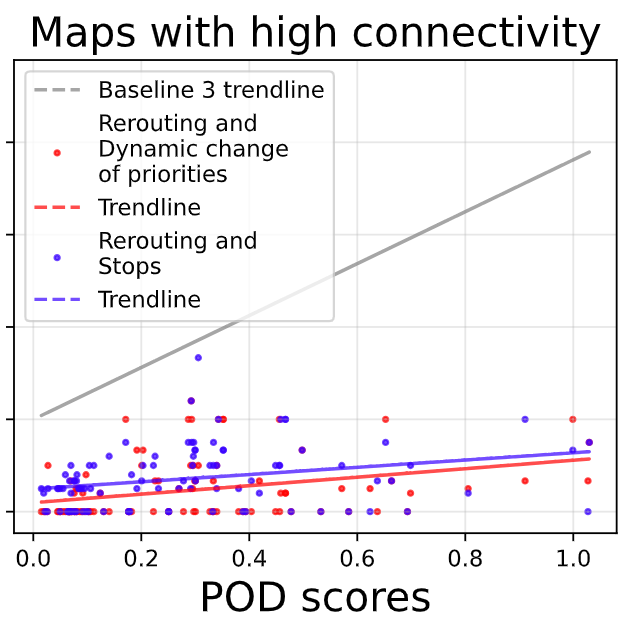}
        \caption{Case III. Scenario 3.1 versus Scenario 3.2.}
        \label{fig:strategy_comparison_32_col}
    \end{subfigure}
    \caption{Strategies comparison in terms of safety (collision rate).}
    \label{fig:str_cmp_safety}
\end{figure*}

Conversely, the strong correlation between POD scores and the outcomes in Baseline scenarios I and III, coupled with the poorest safety outcomes, and the primarily weaker correlations in efficiency across different scenarios where strategies are applied, allows us to draw stronger conclusions about the impact of coordination strategies on system safety (Fig.~\ref{fig:str_cmp_safety}). A trendline for the strategies in use, remaining mostly parallel, which shows the lowest intercept on the Y-axis (indicating collision rates) at the same level of POD scores, will point to the most effective strategy for safety under stable conditions.

Overall, there is no conclusive evidence to show that any coordination strategy outperforms others consistently across all map types and position configurations in relation to the specified traffic rule violations. The dynamic change of priorities appears to be slightly more effective in terms of efficiency (Fig.~\ref{fig:strategy_comparison_1}) and safety (Fig.~\ref{fig:strategy_comparison_1_col}) in Case I, especially on maps with high connectivity. Conversely, implementing stops shows a slight advantage for efficiency in Case III when the MV moves slowly and the POD score increases, indicating denser path overlaps (Figs.~\ref{fig:strategy_comparison_31} and \ref{fig:strategy_comparison_32}). 
The trendlines in safety results show that the performance of AV fleets is more distinctly influenced by the implemented coordination strategies compared to the baseline outcomes. While the trendlines of various strategies largely remain parallel or overlapping, there is a notable difference in performance relative to the baseline trends across all scenarios involving a slow-moving MV (Figs.~\ref{fig:strategy_comparison_31_col} and \ref{fig:strategy_comparison_32_col}). Regardless of the specific coordination strategy employed, the adoption of any such strategy, especially when coupled with rerouting, enhances efficiency and particularly safety outcomes, especially in maps with high connectivity. However, in environments characterized by low connectivity, a decrease in efficiency and safety is often unavoidable due to the lack of rerouting options. Additionally, the presence of two OPs (maps 1, 6, and 10) consistently enhances system performance compared to having just a single OP, as observed in the increase in efficiency. The proximity of AVs to an OP directly impacts operational efficiency. When AVs are positioned closer to an OP, their travel paths are shortened, and path overlaps are reduced, resulting in lower POD scores (Fig.~\ref{fig:pod_scores}) and a higher number of missions completed within each simulation run.

\section{Conclusion}

Key findings of the study suggest that map features unique to underground mining environments have a clearer effect on the efficiency of AV fleets than the coordination strategies applied to these fleets. On the other hand, coordination strategies play a more substantial role in enhancing safety. 
This observation is particularly important in mining environments where there is potential to optimize spatial and operational configurations. 
Intelligent decisions regarding these configurations could improve both the efficiency and safety of operations.

Path Overlap Density (POD) can serve as a predictive metric for evaluating system performance by identifying position configurations within specific maps and coordination strategies that are likely to exhibit higher efficiency and safety. 
This metric’s ability to forecast optimal outcomes is expected to allow data-driven optimization of fleet operations for both specified map features and coordination strategies.
While the POD scores do not directly quantify the number of missions completed or collisions, lower POD scores are associated with a higher likelihood of more completed missions and a reduced collision rate.

The results confirm POD's predictive capability, highlighting its strong association with the number of missions completed and the collision rate across maps with varying levels of connectivity. While POD is designed to be scale-independent and applicable to any number of vehicles, its performance has not yet been systematically assessed across different map scales and varying AV/MV ratios - an interesting direction for future work. The predictive power of this approach should also be further evaluated in future studies, particularly in mixed traffic scenarios that involve different coordination strategies for AV fleets using machine learning models. It is anticipated that these models will more accurately predict the optimal spatial configuration and coordination strategy in each scenario.

\end{document}